Banner appropriate to article type will appear here in typeset article

# Moving contact lines of power-law fluids: How nonlinear fluid rheology drastically alters stress singularity and dynamic wetting behavior

**David Halpern[1] and Hsien-Hung Wei[2]**

[1]Department of Mathematics, University of Alabama, Tuscaloosa AL 35487, USA
[2]2Department of Chemical Engineering, National Cheng Kung University, Tainan 701, Taiwan
**Corresponding author:** Hsien-Hung Wei, hhwei@mail.ncku.edu.tw

Rate-dependent viscosity in power-law fluids significantly affects contact line stress singularities and moving contact line behavior. Contact line forces show more severe divergence for shear-thickening fluids ($n > 1$) or remain finite for shear-thinning fluids ($n < 1$).

This study develops a new theoretical framework beyond the classical De Gennes-Tanner-Cox-Voinov paradigm for power-law fluids, providing unified understanding of advancing contact line behaviors. The framework introduces new dynamic wetting laws where the apparent dynamic contact angle $\theta_d$ depends critically on the characteristic dissipation length $h^* \propto U^{n/(n-1)}$, fundamentally altering its dependence on contact line speed $U$.

For shear-thinning fluids ($n < 1$), $\theta_d \sim (h/h^*)^{(1-n)/3}$, with contact line motion dissipated within $h^*$ extending beyond local wedge height $h$ without requiring microscopic cutoff. In drop spreading, this size-dependent $\theta_d$ varies with spreading radius $R$, yielding $\theta_d \propto U^{3n/(2n+7)}$ consistent with the spreading law $R \propto t^{n/(3n+7)}$.

For shear-thickening fluids, apparent contact line motion is characterized by $\theta_d \sim (h^*/h_m)^{(1-n)/3}$, where dissipation concentrates within $h^*$ much smaller than microscopic liquid height $h_m$. This relates to the Cox-Voinov law $\theta_d \sim \text{Ca}_{\text{eff}}^{1/3}$ using effective capillary number $\text{Ca}_{\text{eff}} = \eta_f U/\gamma$ with microscopic viscosity $\eta_f \propto (U/h_m)^{n-1}$.

Theoretical predictions show good agreement with experiments, emphasizing nonlinear fluid rheology's crucial role in dynamic wetting and spreading.

## 1. Introduction

When a liquid droplet is placed on a substrate, it spreads due to surface tension effects, driven by the three-phase moving contact line at its expanding periphery. In various applications such as coating, painting, printing, and spraying, this local contact line motion plays a crucial role in dynamic wetting, as it directly influences the overall spreading behavior. The motion of a contact line is often characterized by the relationship of the apparent dynamic contact angle $\theta_d$ to the contact line speed $U$. A well-known example is the Cox-Voinov law for complete wetting Newtonian fluids, which states that $\theta_d \propto U^{1/3}$ (Cox 1986; Voinov 1976) . This, in turn, leads to the spreading radius $R$ evolving over time $t$ according to Tanner's law $R \propto t^{1/10}$ (Tanner 1979). A broader discussion on dynamic wetting and spreading, including further developments, can be found in review articles (De Gennes 1985; Bonn *et al.* 2009; Snoeijer & Andreotti 2013).

Many liquids encountered in practical applications, particularly polymer solutions, exhibit non-Newtonian behavior that deviates significantly from constant viscosity assumptions. The power-law model provides a simple framework for describing such fluids, where the apparent viscosity $\eta$ varies with shear rate $\dot{\gamma}$ according to $\eta = k\dot{\gamma}^{n-1}$, with $k$ representing the consistency index and $n$ denoting the flow behavior index. The flow behavior index classifies fluid types: $n < 1$ characterizes shear-thinning (pseudoplastic) behavior, $n > 1$ indicates shear-thickening (dilatant) behavior, and $n = 1$ corresponds to Newtonian fluids. Experimental and theoretical studies have shown that the dependence of $\theta_d$ on $U$ for power-law fluids deviates significantly from the classical Cox-Voinov law (Carré & Eustache 2000; Wang *et al.* 2007*a*; Min *et al.* 2010; Liang *et al.* 2012; Wang & Zhu 2014; Wang *et al.* 2018). Similarly, the spreading kinetics of power law fluids differ from Tanner's law (King 2001; Starov *et al.* 2003; Betelu & Fontelos 2004; Rafaï *et al.* 2004; Wang *et al.* 2007*b*; Min *et al.* 2010). These differences are expected since nonlinear fluid rheology modifies the $\theta_d - U$ relationship, thereby altering the spreading dynamics. However, because surface tension forces are dissipated by viscous forces at different rates, depending on $n$, the impact on the interface profile near the contact line and, consequently, on $\theta_d$ remains unclear.

The effects of shear-thinning and shear-thickening extend beyond drop spreading and capillary wetting; they play a crucial role in various interfacial processes involving moving contact lines. For example, a recent study by Charitatos *et al.* (2020) on dynamic wetting failure due to air entrainment found that shear-thinning fluids tend to delay the onset of wetting failure, whereas shear-thickening fluids accelerate it. This phenomenon can be attributed to changes in effective viscosity—shear-thinning fluids exhibit lower viscosity, increasing the critical substrate speed for wetting failure, while shear-thickening fluids exhibit higher viscosity, reducing this critical speed. To fully understand how air entrainment is triggered, particularly in the more severe wetting failure scenarios observed in shear-thickening fluids, a detailed analysis of contact line motion is essential.

Several dynamic contact angle models for power-law fluids have been proposed (Carré & Eustache 2000; Wang *et al.* 2007*a*; Dandapat & Singh 2010; Liang *et al.* 2012; Wang *et al.* 2018), but they are either somewhat incomplete or lack consistency. For instance, some models suggest that the dynamic contact angles of both shear-thinning and shear-thickening fluids follow the same power-law form with respect to contact line speed, differing only in their coefficients (Wang *et al.* 2007*a*; Liang *et al.* 2012; Wang *et al.* 2018). However, this is unlikely, as the dissipation mechanisms differ between the two cases. In shear-thinning fluids, dissipation primarily occurs in the bulk (Carré & Woehl 2002; Rafaï *et al.* 2004), whereas in shear-thickening fluids, dissipation is concentrated at the contact line (Wang & Zhu 2014). Another unsettled issue is that most of these models fail to explain the spreading kinetics found in other theoretical studies (Betelu & Fontelos 2004; Starov *et al.* 2003). Perhaps the most problematic inconsistency is that, even for the same power-law fluid type, different models predict conflicting dependences on the contact line speed (Min *et al.* 2010).

To address the above issues, it is essential to develop a rigorous hydrodynamic theory for providing a consistent explanation for the contact line motion of both shear-thinning and shear-thickening fluids. To the best of our knowledge, no explicit expression for $\theta_d(U)$ has been directly derived from the hydrodynamic equations governing the contact line motion of power-law fluids. In other words, a Cox-Voinov-like description for the contact lines of power-law fluids is still lacking. Such a description will reveal not only how $\theta_d$ varies precisely with $U$, but also how relevant length scales influence the interface behavior near and away from the contact line. This understanding is crucially dependent on how the nature of the contact line stress singularity changes with fluid rheology, offering plausible microscopic mechanisms (if necessary) to alleviate the singularity (De Gennes 1985; Eggers & Stone 2004; Wei *et al.* 2019). It is worth noting that Wang *et al.* (2007*a*) solved the lubrication equation for the film thickness and obtained the dynamic contact angle relationships for both shear-thinning and shear-thickening fluids. In their expressions, a microscopic cutoff was introduced to remove the contact line singularity. However, as pointed out by Wang & Zhu (2014), such a microscopic cutoff is unnecessary for shear-thinning fluids, as there is no divergence in viscous dissipation which mainly occurs in the bulk (Carré & Woehl 2002; Rafaï *et al.* 2004). For shear-thickening fluids, on the other hand, Wang *et al.* (2007*a*) introduced a microscopic cutoff length $x_m$ to mitigate the much more severe contact line singularity, evaluating the dynamic contact angle at a fixed $x_m$. However, keeping $x_m$ fixed implies that the corresponding contact line microstructure is rate independent, which is generally not the case. Wang & Zhu (2014) developed a scaling theory to address this issue. But a detailed description for the contact line microstructure and how it matches to the macroscopic wedge are still lacking.

Motivated by these challenges, this work aims to develop a rigorous and comprehensive theory to elucidate the behavior of moving contact lines in power-law fluids. We will not only derive the apparent dynamic contact angle relationships for both shear-thinning and shear-thickening fluids in line with their respective stress singularities but also resolve the contact line microstructure by including molecular force effects. Before developing the theory in detail in sections 4 and 5, it is essential to first provide an overview of the relevant concepts in sections 2 and 3. This will offer a clearer understanding of why and how the rheological exponent $n$ influences the characteristics of moving contact lines. Additionally, it will explain why the resulting expressions must take forms fundamentally different from those of the Newtonian counterpart.

## 2. Review of the classical De Gennes-Tanner-Cox-Voinov theory for Newtonian fluids

In this section we consider the advancing motion of a contact line having a small contact angle $\theta_d$ in a wedge geometry (see Figure 1), and start by reviewing some fundamental aspects for the Newtonian case. How the local liquid profile $h(x)$ varies with distance $x$ to the contact line can be described by the well-known Tanner equation, derived using lubrication theory (Tanner 1979):

$$h''' = -\frac{3Ca}{h^2}, \quad (2.1)$$

where $Ca = \eta U/\gamma$ is the capillary number with $\gamma$ being the surface tension. Here primes denote derivatives with respect to $x$. Compared to the more generalized equation for power-law fluids (see (4.1) and its derivation in Appendix A), the form of (2.1) is somewhat special, as its solution takes the form (Bonn *et al.* 2009)

$$h = \ell H(x/\ell). \tag{2.2}$$

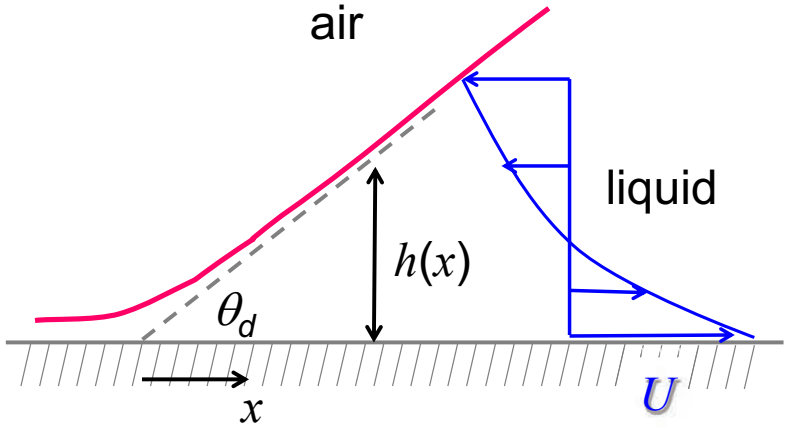

Figure 1. Schematic picture of an advanced contact line motion in the co-moving frame.

Equation (2.2) indicates that the slope $h'$ remains invariant regardless of the characteristic length $\ell$ chosen to scale both $h$ and $x$. This means that if it is taken as a microscopic length, the local contact line dynamics becomes decoupled from the macroscopic droplet motion. Consequently, the apparent dynamic contact angle $\theta_d(x) \equiv h'$ can be determined at $x \gg \ell$. This separation of length scales ensures that the expression for $\theta_d$ becomes universal, taking the following asymptotic form with vanishing curvature at infinity,

$$h'^3(x) = h'^3(\ell) + 9Ca \ln\left(\frac{x}{\ell}\right), \tag{2.3}$$

which is the well-known Cox-Voinov law (Cox 1986; Voinov 1976). As indicated by (2.3), $\theta_d$ exhibits only a weak dependence on $x$ due to the presence of the logarithmic term. This implies that $\theta_d$ is largely insensitive to macroscopic length scales, aligning with the concept of length scale separation. Similarly, the dependence of $\theta_d$ on the microscopic length $\ell$ is also weak in a logarithmic manner, reflecting scale invariance from (2.2). This logarithmic dependence arises from the integration of the diverging viscous stress $\sim \eta U/\theta_d x$ at the contact line within the macroscopic wedge (Huh & Scriven 1971).

Since $\theta_d$ given by (2.3) approximately follows $\theta_d \propto U^{1/3}$ (neglecting the microscopic contact angle $\theta_m \equiv h'(\ell)$), this together with $U = dR/dt$ under the constant volume constraint $R^3\theta_d = \Omega$ renders the spreading radius $R$ to evolve as $R \propto t^{1/10}$, which is the well-known Tanner law (Tanner 1979). Thus, as a direct consequence of the local dynamic contact angle relationship, the global spreading law also inherently reflects the fundamental nature of a moving contact line.

When the spreading liquid becomes a power-law type, the viscosity $\eta$ is no longer constant but depends on the shear rate, leading to a series of changes in the contact line dynamics. Most notably, the viscous term in (2.1) scales as $1/h^\nu$ with $\nu \neq 2$. To preserve scale invariance in the solution, $x$ must be scaled with a different power of $\ell$ than $h$ (Boatto *et al.* 1993). However, this inevitably causes the slope $h'$ to strongly depend on $\ell$, meaning that the local contact line dynamics may no longer decouple from the global drop spreading motion. Therefore, $\theta_d$ and $R$ will no longer obey the Cox-Voinov law (2.2) and Tanner's law. As a result, macroscopic lengths such as drop size or substrate's curvature may explicitly influence $\theta_d$, which has indeed been observed experimentally (Min *et al.* 2010).

We emphasize that such lack of separation between local wetting and global spreading is not exclusive to power-law fluids. Rather, it is a direct consequence of the absence of scale invariance or separation, which can result from any mechanism that modifies the viscous term in (2.1). Another example of this phenomenon can be seen in the spreading of a slipping drop, especially when the slip length is much larger than the drop height (Wei *et al.* 2019).

## 3. Key features of moving contact lines in power-law fluids: scaling analysis

To better illustrate how the characteristics of a contact line are altered in a power-law fluid, it is useful to examine some fundamental aspects and analyze how they are altered by the power-law viscosity

$$\eta = k\dot{s}^{n-1}, \qquad (3.1)$$

where the shear rate $\dot{s} = |\partial u/\partial y|$ measures the strength of the velocity gradient in the transverse direction, $\partial u/\partial y$ ($< 0$ in the frame moving with the contact line, see Figure 1). Building on these aspects, we will then employ a scaling analysis to highlight key features without explicitly solving the hydrodynamic equations.

### 3.1. *Stress singularity versus force singularity: distinct dissipation length scales*

To begin, we recognize that even in a power-law fluid, a stress singularity still persists at a moving contact line due to a velocity jump between the moving air-liquid interface and the no-slip substrate. This can be seen from the fact that the viscous stress $\tau$ with the power index $n > 0$ diverges as the contact line is approached:

$$\tau = \eta \frac{\partial u}{\partial y} = k\left(-\frac{\partial u}{\partial y}\right)^n \sim k\left(\frac{U}{h}\right)^n \to \infty \text{ as } h \to 0. \qquad (3.2)$$

However, the viscous braking force $F_{\text{brake}}$ on the contact line does not necessarily diverge at the contact line as can be seen by integrating (3.2) with respect to $x$. Specifically, with $h \approx \theta_d x$, this force (per unit length) for $n \neq 1$, $F_{\text{brake}}$ behaves as

$$F_{\text{brake}} = k\int_{x_m}^{L}\left(-\frac{\partial u}{\partial y}\right)^n dx \sim k\int_{x_m}^{L}\left(\frac{U}{\theta_d x}\right)^n dx = \frac{k}{1-n}\left(\frac{U}{\theta_d}\right)^n\left[L^{1-n} - x_m^{1-n}\right], \qquad (3.3)$$

where $L$ is the length of the macroscopic wedge and $x_m$ ($\ll L$) is a mesoscopic length bridging the macroscopic and microscopic regions. As evident from (3.3), whether $F_{\text{brake}}$ is dominated by $L$ or $x_m$ critically depends on whether $n < 1$ or $n > 1$. Note that a similar expression to (3.3) can also be derived from the viscous energy dissipation rate over the wedge:

$$T\dot{\Sigma}_w \equiv \int_{x_m}^{L}\int_0^h \eta\left(\frac{\partial u}{\partial y}\right)^2 dy\, dx \sim \frac{k}{1-n}\left(\frac{U}{\theta_d}\right)^{n+1}\left[L^{1-n} - x_m^{1-n}\right], \qquad (3.4)$$

from which (3.3) can be readily recovered using $F_{\text{brake}} \sim T\dot{\Sigma}_w/U$.

For shear-thinning fluids ($n < 1$), $F_{\text{brake}}$ is dominated by the macroscopic $L$ term (Carré & Eustache 2000):

$$F_{\text{brake}}(n < 1) \sim \frac{k}{1-n}\left(\frac{U}{\theta_d}\right)^n L^{1-n}. \tag{3.5}$$

Thus, despite the stress singularity at the contact line, $F_{\text{brake}}$ is actually integrable, making it finite at the contact line regardless of $x_m$. This is qualitatively different from the Newtonian case where $F_{\text{brake}}(n = 1) \sim (\eta U/\theta_d)\ln(L/x_m)$, which diverges logarithmically as $x_m \to 0$ (De Gennes 1985). As viscous dissipation here is no longer concentrated at the contact line but grows with $L$, it suffices to dissipate the local contact line motion through the bulk without needing any microscopic cutoff. As will be seen in section 3.2, this will cause $\theta_d$ to depend on the extent of spreading.

For shear-thickening fluids ($n > 1$), on the contrary, the force (3.3) is mainly controlled by the mesoscopic $x_m$ term:

$$F_{\text{brake}}(n > 1) \sim \frac{k}{1-n}\left(\frac{U}{\theta_d}\right)^n x_m^{1-n}. \tag{3.6}$$

Here $F_{\text{brake}}$ diverges as $x_m \to 0$, indicating a singularity at the contact line. This divergence is of power-law type and thus more severe than the logarithmic divergence seen in the Newtonian case. Therefore, a microscopic cutoff is necessary to prevent the force singularity at the contact line.

Characteristic differences between the $n < 1$ and $n > 1$ cases have been noticed by Liang *et al.* (2012) from an energy dissipation perspective. However, their expression for viscous dissipation energy followed the bulk dissipation mode (3.5) derived by Carré & Eustache (2000) for shear-thinning fluids ($n < 1$) and was extended to the shear-thickening case ($n > 1$) by introducing an additional dissipation factor to account for the effects of stress singularity. As a result, their formulas for $\theta_d$ showed little qualitative difference between these two cases, apart from the dissipation factor. This challenges the notion that their contact line motions are dissipated within different dissipation length scales, as indicated by (3.5) and (3.6). Therefore, their dynamic contact angles must exhibit distinct behaviors, which we demonstrate next.

### 3.2. *Energy dissipation and scaling relationships for $\theta_d$*

Having obtained the distinct viscous force characteristics (3.4) and (3.5), we now balance each with the surface tension force to derive the corresponding scaling relationships for $\theta_d$. This will enable us to reveal essential features of how $\theta_d$ varies with $U$ and the relevant length scales.

The surface tension force that drives the contact line is often expressed as a pulling tension force at the dynamic state relative to the static state with the equilibrium contact angle $\theta_e (< \theta_d)$ (Carré & Eustache 2000):

$$F_s = \gamma(\cos\theta_e - \cos\theta_d). \tag{3.7}$$

For small $\theta_d$, as in the case of complete wetting, (3.7) simplifies to (De Gennes 1985):

$$F_s \approx (\gamma/2)\theta_d^2 + S. \tag{3.8}$$

Here $S = \gamma_{SV} - \gamma_{LS} = \gamma(\cos\theta_e - 1) \approx -(\gamma/2)\theta_e^2$ is the spreading coefficient, which measures the excess interfacial free energy per unit area. The terms $\gamma_{SV}$ and $\gamma_{LS}$ represent the solid-vapor and the liquid-solid interfacial tensions, respectively.

As indicated by (3.8), this surface tension force includes an additional contribution from *S* to account for the wettability of a substrate. In the case of complete wetting, it is well established that for Newtonian fluids, the spreading kinetics described by Tanner's law remains universal, independent of the value of *S* (Bonn *et al.* 2009). This implies that *S* does not enter the dynamic contact angle relationship $\theta_d(U)$, which also explains the universality of the Cox-Voinov law (2.1).

De Gennes (1985) has shown for the Newtonian case that the excess interfacial free energy *S* is virtually consumed by the precursor film ahead of the macroscopic wedge. Here we establish that this also holds for power-law fluids. As shown analytically in Appendix B, the viscous dissipation energy (per unit time) within the film $T\dot{\Sigma}_f$ is nearly equal to the work (per unit time) done by *S*:

$$T\dot{\Sigma}_f \equiv \int_{-d}^{0} \int_{0}^{h} \eta \left(\frac{\partial u}{\partial y}\right)^2 dy\, dx \approx SU, \tag{3.9}$$

where *d* is the width of the "truncated" film with thickness given by $h_e = a(3\gamma/2S)^{1/2}$ with *a* being the molecular length associated with disjoining pressure (De Gennes 1985). The complete burnout of *S* by the precursor film also indicates that the effects of substrate wettability or $\theta_e$ will only enter the contact line microstructure without altering how the surface tension force drives the macroscopic contact line. Previous studies used the full tension force (3.7) to derive dynamic contact angle relationships for small $\theta_d$, often neglecting *S* in (3.8) under the assumption $\theta_e \ll \theta_d$ (Dandapat & Singh 2010; Liang *et al.* 2012). However, if $S \approx -(\gamma/2)\theta_e^2$ is included, this can lead to an overestimation of $\theta_d$ or even fundamentally change its dependence on *U*.

Thus, only the dynamic component $(\gamma/2)\theta_d^2$ in (3.8) will be balanced to the viscous braking force over the wedge. In the following, we demonstrate that the distinct dissipation length scales in (3.5) and (3.6) for the $n < 1$ and $n > 1$ cases will result in fundamentally different characteristics of the corresponding dynamic contact angles.

For shear-thinning fluids ($n < 1$), balancing $(\gamma/2)\theta_d^2$ to (3.5) gives

$$\theta_d^{n+2} \sim \frac{k}{\gamma} U^n L^{1-n}, \tag{3.10a}$$

or expressed in terms of the local wedge height $h \sim L\theta_d$ :

$$\theta_d^3 \sim U^n h^{1-n}. \tag{3.10b}$$

Thus, $\theta_d$ increases with the macroscopic scale *L* or *h* over which the contact motion is mainly dissipated, as indicated by (3.5). In drop spreading problems, *L* corresponds to the spreading radius *R*, making $\theta_d$ depend on the extent of spreading. Given $R = (\Omega/\theta_d)^{1/3}$ from the constant volume constraint $R^3\theta_d = \Omega$, this changes the dependence of $\theta_d$ on *U* in (3.10a) to

$$\theta_d \propto U^{3n(2n+7)}. \tag{3.11}$$

It is worth pointing out that while the mesoscopic length $x_m$ does not enter the apparent contact angle $\theta_d$ at the macroscopic level, it may still affect the inner portion of the contact line as *L* approaches $x_m$ or *h* approaches $h_m$ in (3.10). In other words, although a shear-thinning fluid does not require an additional microscopic cutoff to avoid the stress singularity in the macroscopic contact line motion, it may still admit a meso/micro contact line structure. This situation is similar to that of a strongly slipping moving contact line, where the dynamic contact angle is found to be proportional to $L^{1/2}$ or $h^{1/3}$ (Wei *et al.* 2019), corresponding to the strong shear-thinning $n \to 0$ limit in (3.10a).

As for shear-thickening fluids ($n > 1$), by balancing the tension force $(\gamma/2)\theta_d^2$ to the viscous braking force (3.6), we obtain the following scaling for $\theta_d$:

$$\theta_d^{n+2} \sim \frac{k}{\gamma} U^n x_m^{1-n}, \qquad (3.12a)$$

or, written in terms of the mesoscopic meniscus height $h_m \sim x_m \theta_d$:

$$\theta_d^3 \sim \frac{k}{\gamma} U^n h_m^{1-n}. \qquad (3.12b)$$

In contrast to (3.10), (3.12) now involves the mesoscopic length $x_m$ or height $h_m$, which helps alleviate the contact line singularity due to (3.6). It is important to note that these microscopic length scales are generally not fixed but vary with $U$, depending on the detailed contact line microstructure (De Gennes 1985; Eggers 2005; Wei *et al.* 2019). Typically, $h_m \propto U^{-\alpha}$ where $\alpha > 0$, with its value determined by the nature of the contact line microstructure (Wang & Zhu 2014). This alters the dependence of $\theta_d^3$ on $U$ in (3.12b) to

$$\theta_d^3 \propto U^{n+\alpha(n-1)}, \qquad (3.13)$$

implying that the dynamics of the macroscopic contact line are strongly influenced by the microscopic structure of the contact line.

Note that the above results are derived for the complete wetting scenario where the contribution from the spreading coefficient $S \approx -(\gamma/2)\theta_e^2$ is not included and is entirely dissipated within the precursor film according to (3.9). However, in the case of a partial wetting situation, this contribution becomes significant and can no longer be negligible. By incorporating this term into (3.8) and balancing it with (3.6), (3.13) is modified to

$$\theta_d^{2+n} \sim \theta_d^n \theta_e^2 + \frac{k}{\gamma} U^n x_m^{1-n}. \qquad (3.14)$$

Unlike the complete wetting case, the microscopic cutoff $x_m$ here is a fixed quantity, typically dependent on the equilibrium contact angle $\theta_e$ and the equilibrium film thickness $h_e = a(3\gamma/2S)^{1/2}$ (Eggers 2005). During the early stages of spreading where $\theta_d \gg \theta_e$ and the contact line speed $U$ is high, $\theta_d$ can still be described by (3.13), which can be deduced from (3.14). As spreading progresses and $\theta_d$ approaches $\theta_e$ with a significant reduction in $U$ at later stages, the behavior of $\theta_d$ transitions to that described by (3.14).

## 4. Contact line hydrodynamics

### 4.1. *Modified Tanner equation and characteristic dissipation length*

Assuming small slope $h' \ll 1$ as in the case of complete wetting, we apply the standard lubrication theory to derive the equation governing the contact line motion for power-law fluids (see Appendix A for a detailed derivation). This yields the modified Tanner equation,

$$h''' = -K \frac{U^n}{h^{n+1}}, \qquad (4.1)$$

where $K = (k/\gamma)(2 + 1/n)^n$. The interface has to satisfy the condition that its curvature vanishes at infinity,

$$h''(x \to \infty) = 0. \qquad (4.2)$$

Compared to (2.1) for the Newtonian case, (4.1) can be re-arranged to

$$h''' = -\frac{1}{h^2}\left(\frac{h^*}{h}\right)^{n-1}. \qquad (4.3)$$

Here the factor $(h^*/h)^{n-1}$, with the exponent $(n-1)$ arising from the power-law viscosity (3.1), quantifies the contact line resistance relative to the Newtonian case's $1/h^2$. This resistance factor is characterized by the inherent length scale

$$h^* = (KU^n)^{\frac{1}{n-1}}. \tag{4.4}$$

Consequently, $h^{*n-1}$ in (4.3) essentially serves as an effective capillary number that measures the extent of viscous dissipation:

$$h^{*n-1} = Ca_0^n h_0^{n-1}, \tag{4.5}$$

where $Ca_0 = \eta_0 U/\gamma$ is the capillary number based on the reference fluid viscosity $\eta_0$ and the thickness $h_0 = [(k/\eta_0)(2+1/n)^n]^{1/(n-1)}(\gamma/\eta_0)$ represents the associated length scale.

In view of the above, $h^*$ can be regarded as the characteristic dissipation length. In fact, because $h^* \propto U^{n/(n-1)}$, it actually controls the extent of dissipation depending on the value of $n$. It is evident that $n = 1$ reduces (4.3) to the usual Newtonian case (2.1), where dissipation is independent of length scales. For shear-thinning fluids ($n < 1$), $h^*$ diverges as $U \to 0$, implying that the dissipation extends to the bulk as the contact line advances while its speed gradually decreases. In the extreme shear-thinning limit ($n \to 0$), (4.3) reduces to the equation governing the motion of a strongly slipping contact line (Wei *et al.* 2019) whose dissipation mainly occurs within $h^*$ proportional to the slip length $\lambda$. Conversely, for shear-thickening fluids ($n > 1$), $h^*$ contracts toward the contact line $h = 0$ as $U \to 0$, signifying that the dissipation becomes increasingly localized at the contact line as the wetting progresses. When $n$ is large, the rate of shrinkage of $h^*$ approaches its maximum with $h^* \to U$, corresponding to extremely shear-thickening fluids (Starov *et al.* 2003).

With the aid of $h^*$, the scaling form of (4.3) can immediately reveal different contact line characteristics for shear thinning ($n < 1$) and shear thickening ($n > 1$). Let the local wedge length $x$ scale as $\mathscr{L}$. With the interface slope $h' = \tan\theta \sim h/\mathscr{L}$, (4.3) yields the following scaling for the local interface angle $\theta$:

$$\tan^3\theta \sim \left(\frac{h}{\mathscr{L}}\right)^3 \sim \left(\frac{h^*}{h}\right)^{n-1}. \tag{4.6}$$

For shear-thinning fluids ($n < 1$), $\theta$ follows

$$\tan\theta \sim \left(\frac{h}{h^*}\right)^{(1-n)/3}. \tag{4.7}$$

To achieve a favorable wetting with a small $\theta$, we must have $h^* \gg h$, implying that the contact line's dissipation has to extend beyond the local wedge height $h$, consistent with the bulk dissipation described by (3.5). Equation (4.7) is also consistent with (3.10*b*), showing that the wetting can be enhanced with $\theta \to 0$ as $h \to 0$ due to the drag reduction effect of shear thinning.

In contrast, for shear-thickening fluids ($n > 1$), (4.6) indicates that $\theta \to \pi/2$ as $h \to 0$, implying a complete lack of wetting. In fact, in this case the lubrication equation (4.3) breaks down near the contact line where the interface steepens significantly due to the mounting viscous resistance caused by the more severe contact line singularity. To enable wetting with $\theta \ll 1$ and mitigate the interface's steepening and resistance, it is necessary to introduce a microscopic mechanism operating at a scale $h_m$ much larger than the dissipation length $h^*$ in the vicinity of the contact line. Consequently, if wetting occurs, a well-defined apparent dynamic contact angle can only exist if the interface satisfies $h^*/h_m \ll 1$ in (4.6):

$$\tan\theta \sim \left(\frac{h^*}{h_m}\right)^{(n-1)/3}, \tag{4.8}$$

which is consistent with (3.12*b*).

Regarding the actual solution to (4.1), it is worth revisiting what has been done previously. Wang *et al.* (2007*a*)) analyzed spreading power-law fluid drops and reported that (4.3) admits an analytical solution in a simple power-law form:

$$h = AU^{n/(n+2)} x^{3/(n+2)}, \tag{4.9}$$

with $A > 0$ being the coefficient required to satisfy (4.1). A similar solution form was also identified by Carré & Eustache (2000) for shear-thinning fluids ($n < 1$). In fact, this solution was first obtained by Boatto *et al.* (1993) who investigated various classes of travelling-wave solutions of the evolution equation $h_t + (h^{n+1} h_{xxx})_x = 0$. However, as pointed out by Bertozzi & Pugh (1996), (4.9) is valid only for $n < 1$ but not for $n \geq 1$. Indeed, it can be readily verified that (4.9) does not satisfy (4.1) for $n > 1$, since the left-hand side of (4.1) remains positive while the right-hand side is negative. Moreover, even if (4.9) were assumed in this case, it would result in $h'$ diverging at the contact line ($x = 0$), yielding an apparent contact angle $\theta_d = 90^o$. This is clearly unphysical, as the contact line would not advance at all, not to mention that this contradicts the small slope assumption under which (4.1) is derived.

It is worth noting that due to non-linearity, equation (4.1) can exhibit a variety of solution characteristics, depending on the value of $n$ (Boatto *et al.* 1993). While an exact solution can sometimes exist (Kader *et al.* 2013), it is often not easily converted into a form that facilitates a clear understanding of the underlying physics. Furthermore, even with a simple solution form like (4.9), it is difficult to determine how the dynamic contact angle $\theta_d = h'$ is influenced by the length scales involved in connection to the nature of the contact line singularity. Therefore, an approximate yet more physically oriented approach is needed to solve (4.1), which is presented next.

### 4.2. *New dynamic wetting laws*

By assuming that $h'$ is small here, we will use the slowly varying slope approach to solve equations (4.1) or (4.3) approximately. This approach was first introduced by Voinov (1976, 1977) in his derivation of (2.3) and later extended by Snoeijer (2006) for the large slope scenario. The essence of this approach is that, by assuming the shape of the macroscopic wedge to vary slowly with respect to the distance to the contact line, the dynamic contact angle formula can be derived through a straightforward integration of the equation. The robustness of this approach has been validated by the fact that the numerical error due to this approximation, when compared to the full formula derived by Cox (1986), does not exceed 1% for $\theta_d < 3\pi/4$ (Voinov 1976; Bonn *et al.* 2009). Additionally, it has been demonstrated that this approach can be successfully applied in the modelling of surfactant superspreaders (Wei 2018) and in the theory of slipping moving contact lines (Wei *et al.* 2019).

In the derivation below, following the approach of Voinov (1976, 1977), we use $h$ as the variable to integrate (4.1) under the assumption of slowly varying $h'$. We first re-write $h''' = h' dh''/dh$ in (4.1) and integrate it along with the vanishing curvature condition (4.2). This yields

$$h'' = -KU^n \int_{h_\infty}^{h} \left(\frac{1}{h'}\right) \frac{dx}{h^{n+1}}, \tag{4.10}$$

where $h_\infty \equiv h(x \to \infty)$. Since $h'$ varies slowly, the integral on the right-hand side of (4.10) with $n > 0$ can be approximated as

$$\left(\frac{1}{h'}\right) \int_{h_\infty}^{h} \frac{dx}{h^{n+1}} \approx -\left(\frac{1}{h'}\right) \frac{1}{n} h^{-n}, \quad (4.11)$$

wherein we neglect $h_\infty^{-n}$ in the final result because $h_\infty \gg h$. Substituting (4.11) into (4.10) and re-writing $h'' = h' dh'/dh$, we arrive at

$$(h')^2 \frac{dh'}{dh} = \left(\frac{K}{n}\right) U^n h^{-n}. \quad (4.12)$$

As anticipated, for $n = 1$, integrating (4.12) recovers the Cox-Voinov law (2.3) for the Newtonian case. For power-law fluids with $n \neq 1$, integrating (4.12) yields a new law for the apparent dynamic contact angle $\theta_d \equiv h'(x)$ at $x \gg x_m$:

$$(h')^3 = (h'_m)^3 + \left[\frac{3K}{n(1-n)}\right] U^n \left[h^{1-n} - h_m^{1-n}\right], \quad (4.13)$$

where $h_m$ ($\ll h$) and $h'_m$ ($\ll h'$) are evaluated at a mesoscopic length $x_m$.

As indicated by (4.13), the behavior of $h'$ can be dominated by either the macroscopic $h^{1-n}$ term or the mesoscopic $h_m^{1-n}$ term, depending on whether $n < 1$ or $n > 1$. Which of these is the dominant term arises from distinct dissipation length scales $L$ and $x_m$ for $n < 1$ and $n > 1$, respectively, as described by the braking force behavior in (3.3).

4.2.1. *The shear-thinning case*

For shear-thinning fluids with $n < 1$, (4.13) is dominated by the $h^{1-n}$ term. Neglecting the contributions from $h_m$, (4.13) reduces to

$$(h')^3 = \left[\frac{3K}{n(1-n)}\right] U^n h^{1-n}. \quad (4.14)$$

This confirms the scaling result (3.10b). The corresponding interface profile is then

$$h = \left[\frac{3}{n+2} \left(\frac{3K}{n(1-n)}\right)^{1/3}\right]^{3/(n+2)} U^{n/(n+2)} x^{3/(n+2)}, \quad (4.15)$$

which recovers the exact solution reported previously (Boatto *et al.* 1993; Carré & Eustache 2000; Wang *et al.* 2007a). Re-expressing (4.14) in terms of $x$ using (4.15), we find that the apparent dynamic contact angle $\theta_d \equiv h'(x)$ varies as

$$\theta_d(x) = C_1 U^{n/(n+2)} x^{(1-n)/(n+2)}, \quad (4.16)$$

where $C_1 = [3/(n+2)]^{(1-n)/(n+2)} [3K/n(1-n)]^{1/(n+2)}$. Equation (4.16) also agrees with the result reported by Wang *et al.* (2007a). Note that the above results hold only for $x$ sufficiently far from the microscopic position $x_m$, as the contributions from $h_m$ at that position are neglected in (4.13). Suppose that $\theta_d$ is evaluated at a fixed relative position $r = R - x = (1 - \alpha)R$ with respect to the spreading radius $R$ with $\alpha$ being a constant ($< 1$). Then, $\theta_d$ described by (4.16) not only varies with the extent of the spreading but also is constrained by the drop volume $V = (\pi/2) R^3 \theta_d$. Substituting $x = \alpha R$ and $R = (2V/\pi \theta_d)^{1/3}$ into (4.16), leads $\theta_d$ to vary with $U$ in a manner similar to (3.11):

$$\theta_d = \left[\left(\frac{2V\alpha^3}{\pi}\right)^{1-n} C^{3/(n+2)}\right]^{1/(2n+7)} U^{3n/(2n+7)}. \quad (4.17)$$

Writing $U \sim R/t$ and using $R^3 \theta_d$ = constant again, we can use (4.17) to determine how $\theta_d$ and $R$ vary with time $t$:

$$\theta_d \propto t^{-3n(3n+7)}, \tag{4.18a}$$

$$R \propto t^{n/(3n+7)}. \tag{4.18b}$$

These are consistent with those obtained from the self-similar solution for a spreading shear-thinning drop (King 2001; Starov *et al.* 2003; Betelu & Fontelos 2004). Note that in the strong shear-thinning $n \to 0$ limit, the above results are not applicable since the approximation in (4.11) for the right-hand side of (4.1) is no longer valid. In this limit, the governing equation (4.1) simplifies to

$$h''' = -\frac{K_0}{h}, \tag{4.19}$$

where $K_0 = k/\gamma$ and its inverse is precisely the dissipation length described by (4.4). The same equation form also describes the motion of a strongly slipping contact line with $K_0$ being inversely proportional to the slip length (Wei *et al.* 2019). Following a similar approach as in (4.10)-(4.11), the dynamic contact angle relationship (neglecting $(h'_m)^3$) is found to be

$$(h')^3 = 3K_0 h \ln(h_\infty/h). \tag{4.20}$$

So $(h')^3$ is roughly proportional to $h$, similar to (4.14) with small $n$, albeit it weakly depends on the macroscopic height $h_\infty$ through the logarithmic term. The corresponding interface profile is

$$h \approx (3/2)^{3/2} \left[3K_0 \ln(h_\infty/h)\right]^{1/2} x^{3/2}, \tag{4.21}$$

which approximately follows $x^{3/2}$, resembling (4.15) when $n$ is small. Both (4.20) and (4.21) have also been observed in strongly slipping moving contact lines (Wei *et al.* 2019).

It is worth noting that the viscosity $\eta$ of a shear-thinning fluid does not necessarily approach zero as the shear rate $s$ increases indefinitely, as described in (3.1), particularly near the contact line. Instead, $\eta$ may approach a constant value, known as the infinite-shear viscosity $\eta_\infty$, as $s \to \infty$, as described by the Carreau mode (Bird 1976; Carreau *et al.* 1979). In this case, the contact line structure can be divided into three regions: an inner region that behaves like a Newtonian liquid with viscosity $\eta_\infty$, an outer region that exhibits shear-thinning behavior with a higher viscosity, and an intermediate region that transitions between the two (Charitatos *et al.* 2020).

In fact, the behavior of such Carreau shear thinning fluids is somewhat similar to polymeric liquids where viscosity is significantly reduced near a substrate due to the depletion layer (De Gennes 1985). This results in an apparent slip length $\lambda$ near the contact line where viscosity decreases from the bulk value $\eta_0$ to $\eta_\infty$. As the shear stress near the contact line behaves as

$$\tau = \eta \frac{\partial u}{\partial y} \sim \eta_0 \frac{U}{h+\lambda} \sim \eta_\infty \frac{U}{h}, \tag{4.22}$$

the slip length $\lambda$ can be related to the viscosity reduction ratio through

$$\frac{h}{\lambda} \sim \frac{\eta_0}{\eta_\infty} - 1, \tag{4.23}$$

establishing a connection between shear thinning effects and slip effects.

### 4.2.2. *The shear-thickening case*

For shear-thickening fluids with $n > 1$, we re-write (4.13) in terms of the characteristic dissipation length $h^*$ defined by (4.4):

$$(h')^3 = (h'_m)^3 + \left(\frac{3}{n(n-1)}\right)\left[\left(\frac{h^*}{h_m}\right)^{n-1} - \left(\frac{h^*}{h}\right)^{n-1}\right]. \tag{4.24}$$

Equation (4.24) shows that the slope $h'(x)$ increases monotonically with $h$ as $h$ grows toward the bulk fluid. As $h \to \infty$ when $x \to \infty$, (4.24) approaches a constant value, representing the apparent dynamic contact angle $\theta_d \equiv \tan^{-1}(h'(x \to \infty))$, given by

$$h'(x \to \infty) = \left[(h'_m)^3 + \left(\frac{3}{n(n-1)}\right)\left(\frac{h^*}{h_m}\right)^{n-1}\right]^{1/3}, \tag{4.25}$$

suggesting that wetting under small values of $\theta_d$ would become more favorable when $h^* < h_m$ (see Figure 2a). Since $h^* = (KU^n)^{1/(n-1)}$ according to (4.3), this implies that favorable wetting can only be achieved when the speed $U$ is below a certain critical value, $U_c$:

$$U_c = \left(K^{-1}h_m^{n-1}\right)^{1/n}. \tag{4.26}$$

In this case, assuming $(h'_m)^3$ is negligible, and (4.25) simplifies to

$$h'(x \to \infty) \approx C_2\left(\frac{k}{\gamma}\right)^{1/3} U^{n/3} h_m^{(1-n)/3}, labeleq : hpatinfapprox \tag{4.27}$$

which is (3.12*b*), where $C_2 = [3(2 + 1/n)^{n/n(n-1)}]^{1/3}$.

On the other hand, if $U > U_c$, this means $h^* > h_m$, which may cause the dynamic contact angle $\theta_d$ to no longer be small and even approach $90^o$ (see Figure 2). In this case, wetting becomes less favorable. Note from (4.24) that $h'(x)$ will be smaller than $h'_m \equiv h'(x_m)$ when $h < h_m$ and continue to decrease as $h$ approaches the contact line. If there is a pre-wetted film ahead of the macroscopic contact line, this will allow $h'(x)$ to continuously decrease towards zero, with $h' = 0$ at some $h < h_m$ or $x < x_m$. Since $h'$ cannot be further decreased to a negative value below this point, an additional mechanism is required to ensure that this zero-slope location also acts as an inflection point $h'' = 0$. Below this point, $h'$ must necessarily increase as it moves towards the microscopic contact line. Given that $h$ is very small in this region, this inflection point and subsequent film thinning below it can only arise from disjoining pressure effects. For the potentially large $\theta_d$ situation when $h^* > h_m$, in particular, the presence of a precursor film due to disjoining pressure would become even more essential to facilitate the forward movement of the contact line. In any case, the thickness of the precursor film, which determines $h_m$, and the extent of its spread will be governed by the detailed profile of the film, which is examined in the next section.

## 5. Precursor film in power-law fluids

Ahead of the macroscopic contact line, a precursor film (typically 10-100nm thick) often forms due to van der Waals disjoining pressure (Bonn *et al.* 2009):

$$\Pi = \gamma\frac{a^2}{h^3}, \tag{5.1}$$

(a)

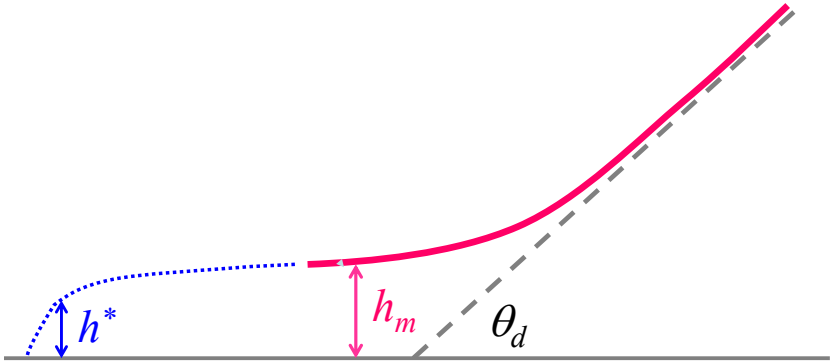

(b)

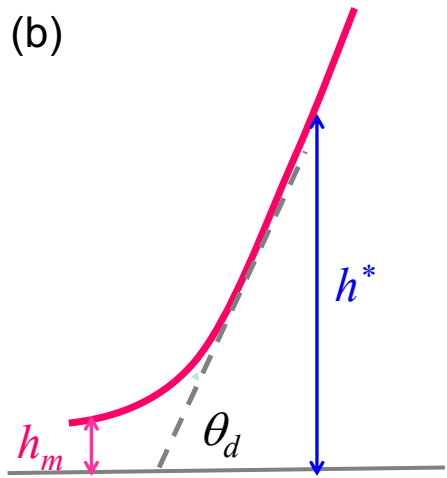

Figure 2. Schematic illustrations showing how the contact line structure for shear-thickening fluids ($n > 1$) is influenced by the relative size of the dissipation length $h^*$ compared to the microscopic cutoff $h_m$. (a) $h^* \ll h_m$. (b) $h_m \ll h^*$.

where $a = (A/6\pi\gamma)^{1/2}$ is the molecular length which is typically an order of 1 Å, and $A$ is the Hamaker constant.

The equation governing the precursor film motion can be derived by adding the disjoining pressure term $-\Pi$ from (5.1) to the capillary pressure $-\gamma h''$, modifying (4.1) to

$$h''' = -K\frac{U^n}{h^{n+1}} + \frac{3a^2 h'}{h^4}. \tag{5.2}$$

We first determine the scales for the precursor film thickness $h_f$ and its lateral extent $\ell$ by balancing the terms in (5.2). This can be more conveniently done by re-writing (5.2) into the form analogous to the Newtonian case:

$$h''' = -\frac{\mathscr{C}}{h^2} + \frac{3a^2 h'}{h^4}, \tag{5.3}$$

where $\mathscr{C} = KU^n/h^{n-1}$ is the effective capillary number that varies with $h$. We begin with $h_f \sim \mathscr{C}\ell_f^3$ by balancing the surface tension term $h''' \sim h_f/\ell_f^3$ with the viscous term $-\mathscr{C}/h^2 \sim \mathscr{C}/h_f^2$. Similarly, balancing the disjoining pressure term $3a^2 h'/h^4 \sim a^2 \ell_f^{-1}/h_f^3$ with the surface tension term $h_f/\ell_f^3$ yields $h_f^2 \sim a\ell_f$. Combining these two relations, we obtain $h_f$ and $\ell_f$ in terms of $\mathscr{C}$ (Bonn *et al.* 2009):

$$h_f \sim a\mathscr{C}^{-1/3}, \quad \ell_f \sim a\mathscr{C}^{-2/3}. \tag{5.4a,b}$$

Re-writing $\mathscr{C} = Ca_f(a/h_f)^{n-1}$ in terms of $h_f$ and the film capillary number $Ca_f = KU^n/a^{n-1}$, we derive the following scaling relations for $h_f$ and from (5.4):

$$h_f \sim aCa_f^{-1/(4-n)}, \quad \ell_f \sim aCa_f^{-2/(4-n)}, \tag{5.5a,b}$$

To solve for the film profile using (5.2), we non-dimensionalize (5.2) using $\phi = h/h_f$ and $\zeta = x/\ell_f$ based on (5.5):

$$\phi''' = -\frac{1}{\phi^{n+1}} + \frac{3\phi'}{\phi^4}, \tag{5.6}$$

where primes denote derivatives with respect to $\zeta$. For numerical calculations, it is more convenient to utilize the asymptotic solution $\phi_{-\infty}$ as $\zeta \to -\infty$, away from the macroscopic contact line. This solution is obtained by balancing the driving viscous $\phi^{-(n+1)}$ term with the disjoining pressure term $3\phi'/\phi^4$, since the film profile in this region is nearly flat (except near the tip) and the surface tension $\phi'''$ term can be considered small compared to the other two terms. This yields

$$\phi_{-\infty} = \left(\frac{3}{(2-n)(-\zeta)}\right)^{\frac{1}{2-n}}. \tag{5.7}$$

The correction to the asymptotic solution (5.7) due to surface tension, denoted as $\psi$, is a static adjustment. It is obtained by balancing the perturbed disjoining pressure term $3\psi'/\phi_{-\infty}^4$ with the surface tension term $\psi'''$:

$$\psi''' = \frac{3\psi'}{\phi_{-\infty}^4}. \tag{5.8}$$

Following Hervet & De Gennes (1984) and Eggers & Stone (2004), the approximate solution to (5.8) can be determined as follows using the WKB method (see Appendix C for detailed derivation):

$$\psi = g_0^{-3/2} \exp\left(\int g_0 d\zeta\right), \tag{5.9a}$$

where

$$g_0 = \frac{\sqrt{3}}{\phi_{-\infty}^2}. \tag{5.9b}$$

The more accurate asymptotic film profile can therefore be constructed by adding (5.8) to (5.7):

$$\phi = \phi_{-\infty} + \Lambda\psi, \tag{5.10}$$

with $\Lambda$ being the adjustable coefficient. For a given value of $\Lambda$, we use (5.10) to specify the values of $\phi$, $\phi'$ and $\phi''$ at $\zeta = \zeta_0 < 0$, which serve as the initial conditions to integrate (5.6) numerically towards $\zeta > 0$. The values of $\Lambda$ and $\zeta_0$ are selected such that the solution asymptotically satisfies the zero curvature condition at infinity, $\phi''(\zeta \to \infty) \to 0$. For a given value of $n$, we find that the value of $\Lambda$ is not too sensitive to the choice of $\zeta_0$ as long as $|\zeta_0|$ is sufficiently large. For $n = 1.1, 1.2, 1.3,$ and $1.4$, the respective values of $\Lambda$ are 3.407, 1.442, 0.512, and 0.073.

Figure 3 displays the computed interface profiles for these $n$ values (scaled by the molecular length $a$) with the cutoff $h_m$ taken as the computed interface height at $x = 0$. These profiles exhibit excellent agreement with the asymptotic wedge profiles corresponding to the apparent contact angle given in (4.25). This result confirms the scaling relation (5.5a) for the film thickness, indicating that the cutoff scales as

$$h_m \sim aCa_f^{-1/(4-n)}. \tag{5.11}$$

We now revisit the relationship (4.25) for the apparent dynamic contact angle $\theta_d$ by re-writing $h^* = (KU^n)^{1/(n-1)}$ from (4.4) in terms of $Ca_f = KU^n/a^{n-1}$:

$$\theta_d \equiv h'(x \to \infty) \sim \left(\frac{h^*}{h_m}\right)^{(n-1)/3} = \left(\frac{a}{h_m}\right)^{(n-1)/3} Ca_f^{1/3}. \tag{5.12}$$

Here, the capillary number is defined as

$$Ca_f = \eta_a \frac{U}{\gamma}, \tag{5.13a}$$

where $\eta_a$ is the characteristic viscosity scale over the molecular length $a$, given by

$$\eta_a = (2 + n^{-1})^n k \left(\frac{U}{a}\right)^{n-1}. \tag{5.13b}$$

Using this, the right-hand side of (5.12) can be recast in terms of the effective capillary number defined by

$$Ca_{\text{eff}} = \eta_f \frac{U}{\gamma}, \tag{5.14a}$$

where $\eta_f$ is the viscosity scale based on the local shear rate $U/h_m$ across the microscopic cutoff length $h_m$:

$$\eta_f = (2 + n^{-1})^n k \left(\frac{U}{h_m}\right)^{n-1}. \tag{5.14b}$$

Substituting (5.14) into (5.12), we recover the familiar scaling for Newtonian fluids:

$$\theta_d = Ca_{\text{eff}}^{1/3}. \tag{5.15}$$

The result above is significant because it elegantly extends classical theories of dynamic wetting — originally developed for Newtonian fluids — to non-Newtonian, power-law fluids, while preserving a familiar scaling structure. The key implications are as follows. First, $Ca_{\text{eff}}$ adapts the classical capillary number $Ca = \eta U/\gamma$ for non-Newtonian fluids by replacing the constant viscosity $\eta$ with a microscopically scaled viscosity $\eta_f$ based on the local film thickness $h_m$. As a result, the apparent contact angle $\theta_d$ captures this intricate interplay between shear-dependent viscosity and film thickness— revealing how microscale physics governs macroscale observables. More importantly, this reformulation leads to a universal scaling law (5.15) that is analogous to the Cox–Voinov law for Newtonian fluids, demonstrating that even in the presence of complex rheology, dynamic wetting behavior maintains a simple and interpretable structure.

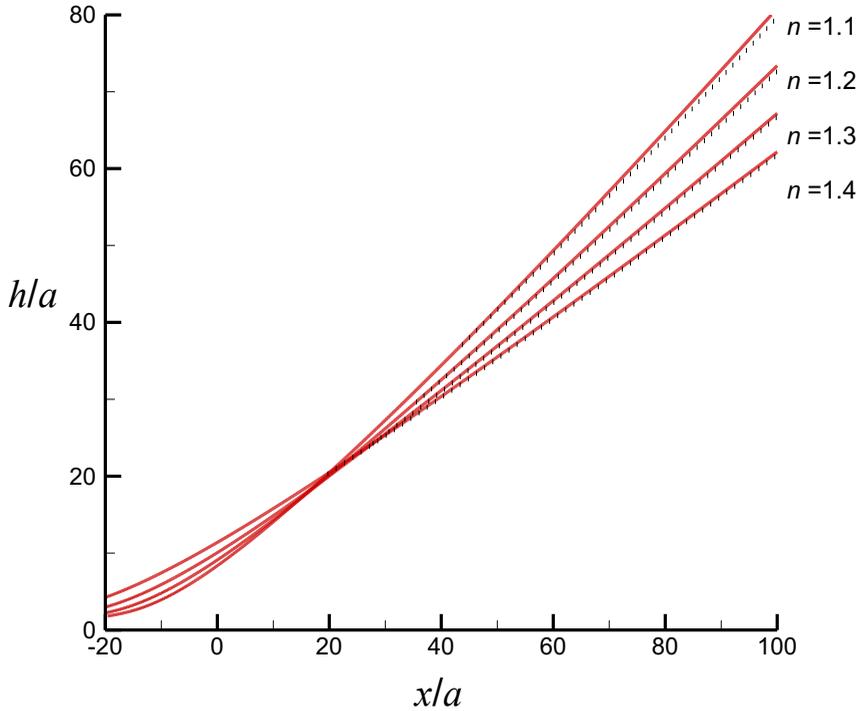

Figure 3. Computed interface profiles (red solid lines) from numerical integration of (5.6) for shear-thickening fluids at different values of $n > 1$. The results perfectly match the corresponding profiles (black dotted lines) in the outer wedge (macroscopic contact angle) region according to (4.25). $Ca_f = KU^n/a^{n-1} = 0.1$.

Along the above lines, the microscopic viscosity from (5.14b), because of $h_m$ from (5.11), increases as a positive power of the contact line speed $U$ according to

$$\eta_f \propto U^{4(n-1)/(4-n)}, \tag{5.16}$$

indicating that for a given value of $U$, $\eta_f$ can be slightly boosted by increasing $n$. Hence, $Ca_{\text{eff}}$ also increases with $U$ in a similar manner as

$$Ca_{\text{eff}} \propto U^{3n/(4-n)}. \tag{5.17}$$

It follows therefore that the apparent dynamic angle from (5.15) varies with $U$ as

$$\theta_d \sim Ca_f^{1/(4-n)} \propto U^{n/(4-n)}, \tag{5.18}$$

which is consistent with the result reported by Wang & Zhu (2014).

## 6. Comparison with experiments

Having established a comprehensive theoretical framework, we now compare our predictions with experimental observations. Figure 4 presents the measured dependence of the dynamic contact angle $\theta_d$ on the spreading speed $U$ from drop spreading experiments (Carré & Eustache 2000; Wang *et al.* 2007*a*) using shear-thinning fluids ($n < 1$). The data are presented on a log-log scale, allowing us to identify the power-law relationship $\theta_d \sim U^m$ from the slope $m$. For the rheological exponent $n = 0.4378$ and $0.79$, the measured power values are $m = 0.16$ and $0.26$, respectively, which are in excellent agreement with the corresponding theoretical values $3n/(2n + 7) = 0.17$ and $0.27$, as given by (4.17). Wang *et al.* (2007*a*) predicted $m = n/(n + 2) = 0.18$ and $0.28$ according to (3.12*a*), which are slightly higher than ours and more applicable to partial wetting situations under which $x_m$ is fixed.

It is important to note that we also recover the same power expression $m = n/(n + 2)$ from the original version of the dynamic contact angle relationship (3.14), as in Wang *et al.*'s formulation. However, this relationship indicates that $\theta_d$ varies with the distance to the contact line, $x$, which is on the order of a macroscopic length $L$. In the context of drop spreading, $L$ is typically taken to be the drop spreading radius $R$, which is not an independent variable but must vary inversely with $\theta_d$ due to the constant drop volume constraint. Taking this interdependence into account leads to the revised scaling law (4.17), yielding a different power-law form: $m = 3n/(2n + 7)$. This contrasts sharply with Wang *et al.* (2007*a*), who treated $x$ as a fixed quantity.

Under the constant drop volume constraint $R^3 \theta_d =$ constant again, the dynamic contact angle relationship $\theta_d \sim U^m \sim (R/t)^m$ also implies a spreading law $R \sim t^\nu$ with the spreading exponent $\nu = m/(m + 3)$, where $t$ is time. Substituting $m = 3n/(2n + 7)$ from our derived $\theta_d - U$ relationship (4.18*a*) yields $\nu = n/(3n + 7)$ as in (4.18*b*), in agreement with the result from the self-similar solution (King 2001; Starov *et al.* 2003; Betelu & Fontelos 2004). Figures 5a and 5b present the log-log plots of $R$ against $t$ using the data measured by Rafaï *et al.* (2004) and Wang *et al.* (2007*b*), from which the spreading exponent $\nu$ is extracted from the slopes. Figure 5c shows that the measured $\nu$ values for various $n$ are in reasonable agreement with the theoretical prediction $\nu = n/(3n + 7)$. While this comparison has been made previously (Wang *et al.* 2007*b*), we include it here to highlight the fact that the spreading exponent $\nu = n/(3n + 7)$ directly follows the contact angle exponent $m = 3n/(2n + 7)$ - a connection not explicitly made until this work. It is also worth noting that the alternative expression $\nu = n/(4n + 6)$ derived from $m = n/(n + 2)$ as proposed by Wang *et al.* (2007*a*) also captures experimental trends, although it is not consistent with the self-similar-solution result.

For shear-thickening fluids ($n > 1$), Figure 6 shows the log-log plots of $\theta_d$ against $U$ using the data measured by Wang *et al.* (2007*a*). For $n = 1.3031$ under complete wetting condition shown in Figure 6a, the measured slope is $0.39$, which is reasonably well captured by our prediction $m = n/(4 − n) ≈ 0.48$ from (5.18). For $n = 1.7282$ for a partially wetting drop, Figure 6b shows that $θ_d$ increases slowly with $U$ at slow speeds, but rises more rapidly at higher $U$, consistent with the behavior described by (3.14). In particular, during the early stage of wetting where $U$ is high and $θ_d$ is expected to follow the scaling in (3.12a), the measured slope is 0.49, which is reasonably close to the theoretical value $m = n/(n + 2) ≈ 0.46$ according to (3.12a) with $x_m$ being kept fixed.

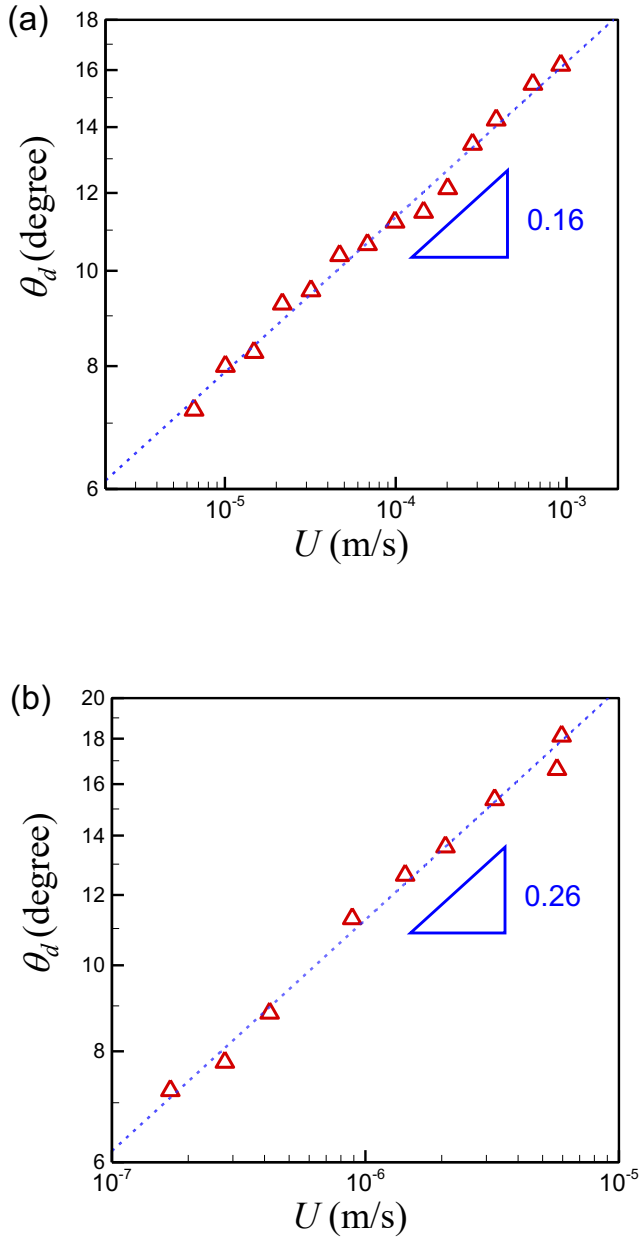

Figure 4. Comparison between experimental data (symbols) and theory predictions for the measured relationship between dynamic contact angle ($\theta_d$) and the wetting speed ($U$) (in log-log scale) in shear-thinning fluids ($n < 1$). (a) In the experiment by Carré & Eustache (2000) with $n = 0.4378$, the best fit (dashed line) is $\theta_d = 48.406 U^{0.1576}$, yielding the slope $\approx 0.16$, in good agreement with our theoretical prediction $3n/(2n+7) = 0.17$ according to (4.17). (b) In the experiment conducted by Wang *et al.* (2007a) with $n = 0.79$, the best fit curve (dashed line) is given by $\theta_d = 412.09 U^{0.2606}$, resulting in the slope $\approx 0.26$, which closely matches with our theoretical prediction $3n/(2n+7) = 0.27$.

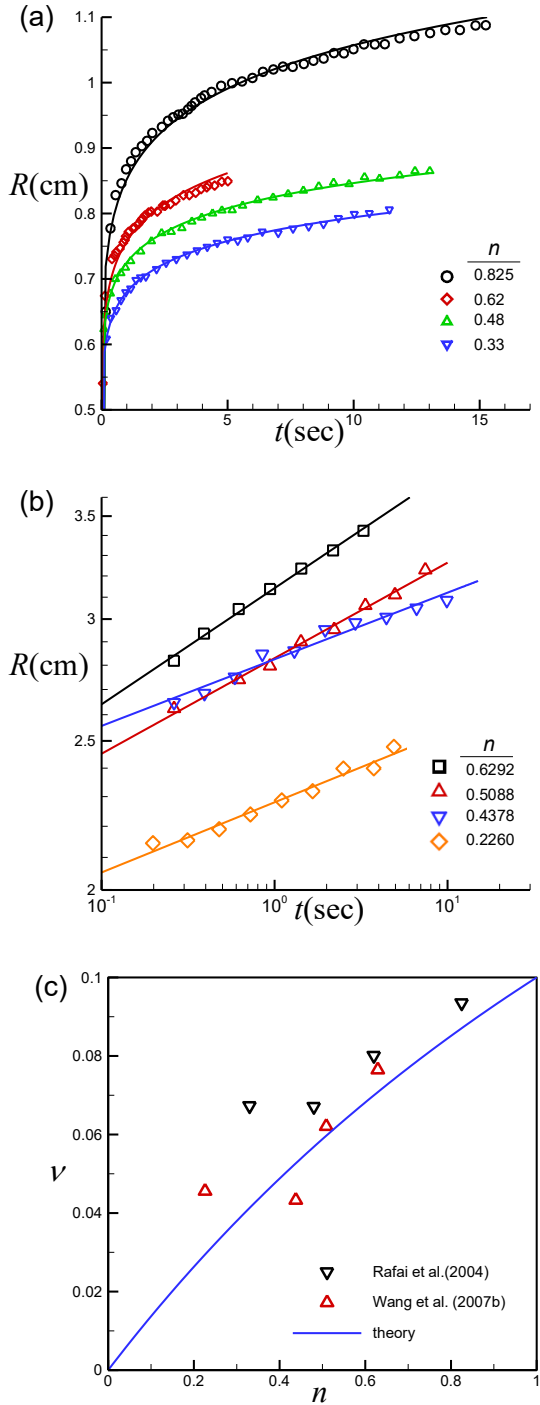

Figure 5. Plots of spreading radius, $R$, versus time, $t$, for shear-thinning fluids ($n < 1$) obtained from drop spreading experiments by (a) Rafaï *et al.* (2004) and (b) Wang *et al.* (2007b). Solid lines represent best fits using $R = At^\nu$ to determine the spreading exponent $\nu$. (c) Comparison between experiment and theory: plot of $\nu$ versus the rheology exponent $n$. The trend of the measured $\nu$ values from (a) and (b) is reasonably captured by the theoretical prediction $\nu = n/(3n + 7)$.

## 7. Concluding remarks

In this work, we theoretically analyze moving contact lines in power-law fluids, demonstrating that their characteristics are intimately linked to the nature of the stress singularity at the contact line, depending on whether the fluid is shear-thinning ($n < 1$) or shear-thickening ($n > 1$). Specifically, the behavior of a contact line in a power-law fluid is highly sensitive to the degree of the contact line stress singularity compared to Newtonian fluids. It is well known that in Newtonian fluids the contact line stress singularity behaves as $1/x$ as the distance $x$ approaches zero at the contact line, leading to a logarithmic divergence in the contact line force. For shear-thinning fluids ($n < 1$), the contact line stress singularity is relatively mild, making it integrable. As a result, the contact line force remains finite without the need of a microscopic cutoff. In contrast, for shear-thickening fluids ($n > 1$), the contact line stress singularity becomes much more severe, leading to an even stronger divergence of the contact line force. In this case, a microscopic cutoff is required to prevent such divergence.

We also develop a rigorous dynamic contact line theory that not only reveals the above features but also derives the dynamic contact relationships, extending the classical De Gennes-Tanner-Cox-Voinov framework for Newtonian fluids. This theory shows that the contact line characteristics depend crucially on the inherent dissipation length $h^*$, which controls the extent of viscous dissipation in power-law fluids. This, in turn, strongly influences how the dynamic contact angle $θ_d$ varies with both macroscopic and microscopic length scales, in addition to its dependence on the contact line speed $U$. For shear-thinning fluids ($n < 1$), $h^*$ can be extended to the bulk, leading $θ_d \propto U^{n/(n+2)} R^{(1-n)/(n+2)}$ to further depend on the extent of spreading, $R$. This leads the dynamic contact angle relationship to be $θ_d \propto U^{3n/(2n+7)}$, as also observed from experiments (Carré & Eustache 2000; Wang *et al.* 2007a). The corresponding spreading law becomes $R \propto t^{n/(3n+7)}$, where $t$ is time. This spreading law is not only consistent with the self-similar solution derived previously (Betelu & Fontelos 2004) but also agrees with experimental observations (Rafaï *et al.* 2004; Wang *et al.* 2007b).

For shear-thickening fluids ($n > 1$), dissipation is concentrated within $h^*$ much smaller than the microscopic liquid height $h_m$ near the contact line. We also consider the precursor film ahead of the macroscopic contact line and determine its profile by including disjoining pressure effects. This provides a precise measure for the required microscopic cutoff $h_m \propto U^{-n/(4-n)}$ in the apparent dynamic contact angle relationship (4.25), leading to $θ_d \propto U^{3n/(4-n)}$. We also show that the $θ_d$–$U$ relationship can be re-expressed as the familiar Cox–Voinov law $θ_d \sim Ca_{\text{eff}}^{1/3}$ in terms of the effective capillary number $Ca_{\text{eff}} = η_f U/γ$ ($γ$ being surface tension) with the microscopic viscosity $η_f \propto (U/h_m)^{n-1}$ based on the local shear rate $U/h_m$ across $h_m$. This enables one to use the classical De Gennes–Tanner–Cox–Voinov framework for Newtonian fluids to power-law fluids, provided the effective viscosity is defined locally in terms of the microscopic film thickness and shear rate. Our results thus bridge the gap between the classical theory and non-Newtonian wetting dynamics, demonstrating that despite the complex rheology of power-law fluids, the dynamic contact angle retains the same qualitative scaling with the effective capillary number. This reinterpretation suggests that, by properly accounting for shear-thinning or thickening effects through $Ca_{\text{eff}}$, one can extend established wetting models to a wider range of complex fluids without fundamentally altering the underlying scaling law. This seems to be consistent with a recent study by Mobaseri *et al.* (2025) who developed a universal framework for predicting the maximum spreading diameter of non-Newtonian drops upon impact by identifying characteristic shear rates through energy budget analysis, enabling mapping to equivalent Newtonian drop behavior and providing design guidelines for controlling impact dynamics.

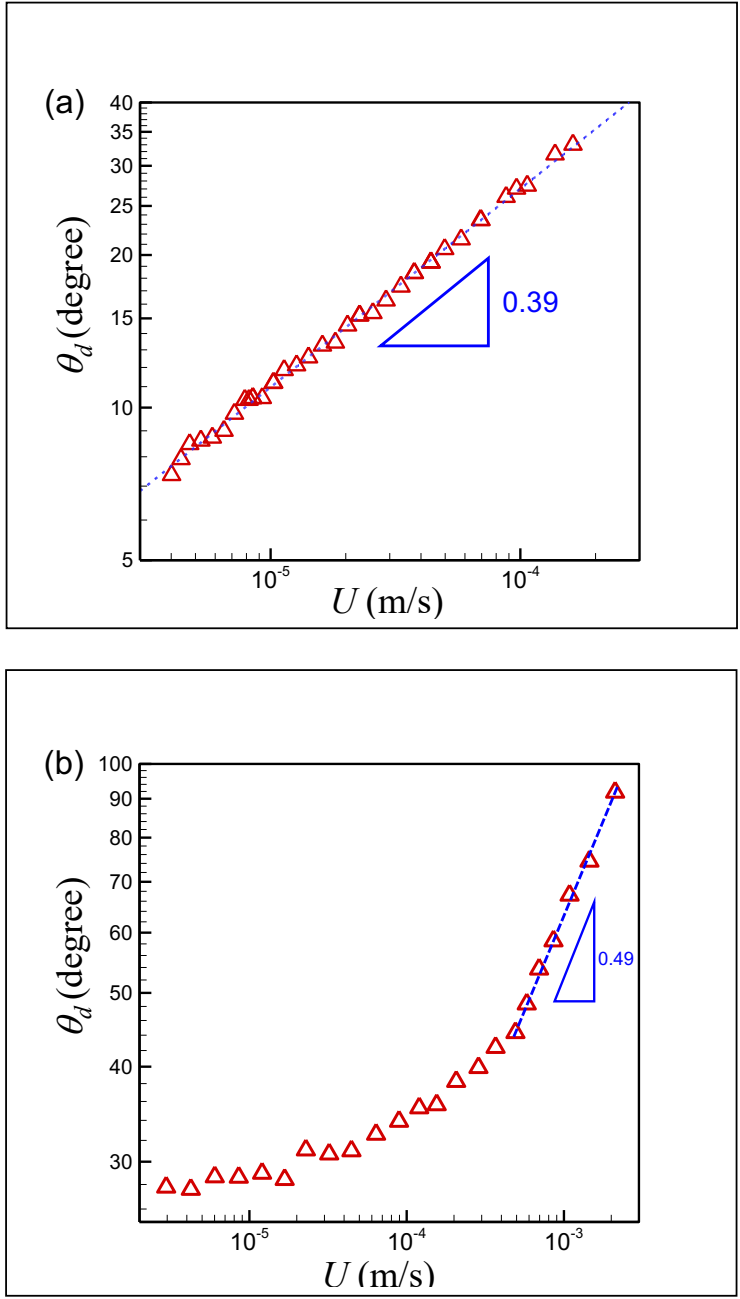

Figure 6. Comparison between experimental data (symbols) of Wang *et al.* (2007*a*) and theory predictions for the measured relationship between dynamic contact angle, $\theta_d$, and the wetting speed, $U$, (in log-log scale) in shear-thickening fluids ($n > 1$). (a) displays the result for $n = 1.3031$ under the complete wetting condition. The best fit curve (dashed line) is given by $\theta_d = 997.21 U^{0.3917}$, yielding the slope $\approx 0.39$, which is reasonably captured by our theoretical prediction $n/(4-n) = 0.48$ according to (5.18). (b) presents the result for a partially wetting drop with $n = 1.7282$. The slope of the data trend in the early stage of spreading is about 0.49, which is reasonably close to our theoretical prediction $n/(n+2) = 0.48$ according to (3.12*a*) with $x_m$ being kept fixed.

**Acknowledgements.** We acknowledge the support from the National Science and Technology Council of Taiwan. HHW also thanks T.-I. Lin for her assistance on the preparation of this manuscript.

**Declaration of interests.** The authors report no conflict of interest.

## Appendix A. Derivation of (4.1) using the lubrication theory

The motion of a contact line is most conveniently described in a reference frame moving with it at speed $U$, as shown in Figure 1. Assuming a small slope, the lubrication theory can be applied, where the flow field ($u$) is predominantly in the horizontal direction ($x$) but varies vertically across the liquid (in $y$). The equation governing the fluid motion is approximately

$$\frac{\partial p}{\partial x} = \frac{\partial}{\partial y}\left(\eta \frac{\partial u}{\partial y}\right). \tag{A 1}$$

Here $p$ is the pressure, and $\eta$ is the viscosity which varies with the shear rate ($-\partial u/\partial y$) ($> 0$) in a power-law manner according to

$$\eta = k\left(-\frac{\partial y}{\partial y}\right)^{n-1}, \tag{A 2}$$

with $k$ being the consistency constant.

The velocity profile can be obtained by solving (A 1) with the no-slip boundary condition $u = U$ at the solid substrate $y = 0$ and the free-stress condition $\partial u/\partial y = 0$ at the air-liquid interface $y = h(x)$, yielding

$$u = \left(1 + \frac{1}{n}\right)^{-1}\left(\frac{1}{k}\frac{\partial p}{\partial x}\right)^{\frac{1}{n}}\left[(h-y)^{1+\frac{1}{n}} - h^{1+\frac{1}{n}}\right] + U. \tag{A 3}$$

Further applying the zero-flow rate condition across the liquid,

$$\int_0^h u\, dy = 0, \tag{A 4}$$

we can relate the pressure gradient to the contact speed as

$$\frac{\partial p}{\partial x} = k\left(2 + \frac{1}{n}\right)^n \frac{U^n}{h^{1+\frac{1}{n}}}. \tag{A 5}$$

By replacing the pressure above with the Laplace pressure at the interface due to surface tension $\gamma$,

$$p = -\gamma \frac{\partial^2 h}{\partial x^2}, \tag{A 6}$$

the evolution equation for the liquid thickness $h(x)$, given by equation (4.1), can then be derived.

## Appendix B. Derivation of (3.9) for viscous dissipation energy

We start with the formula for the viscous dissipation energy within the film:

$$T\dot{\Sigma}_f = \int_{-d}^{0}\int_{0}^{h} \eta \left(\frac{\partial u}{\partial y}\right)^2 dy\, dx, \tag{B 1}$$

with the viscosity $\eta$ being given by (A 2). From (A 3), we can write the shear rate $(-\partial u/\partial y)$ as

$$\left(-\frac{\partial u}{\partial y}\right) = \left[\left(2 + \frac{1}{n}\right)^n \frac{U^n}{h^{n+1}}(h-y)\right]^{\frac{1}{n}}. \tag{B 2}$$

With (B 2) and (A 2), the integrand in (B 1) is

$$\eta \left(\frac{\partial u}{\partial y}\right)^2 = k\left|\frac{\partial u}{\partial y}\right|^{n+1} = k\left(2+\frac{1}{n}\right)^{n+1} \frac{U^{n+1}}{h^{(n+1)(1+1/n)}}(h-y)^{1+1/n}. \tag{B 3}$$

Using (B 3), we first carry out the integral over $y$, leading (B 1) to

$$T\dot{\Sigma}_f = k\left(2+\frac{1}{n}\right)^n U \int_d^0 \frac{U^n}{h^n}\,dx. \tag{B 4}$$

Making use of (5.2) by letting the driving viscous term be mainly balanced by the disjoining pressure term because the surface tension $h'''$ term plays a minor role in the film, the integrand in (B 1) can be approximated as

$$\frac{U^n}{h^n} \approx \left(2+\frac{1}{n}\right)^{-n}\left(\frac{a^2\gamma}{k}\right)\frac{3h'}{h^2}, \tag{B 5}$$

With the above, (B 4) becomes

$$T\dot{\Sigma}_f \approx Ua^2\gamma \int_{-d}^{0} \frac{3h'}{h^3}\,dx. \tag{B 6}$$

The integral can be evaluated as

$$\int_{-d}^{0} \frac{3h'}{h^3}\,dx = \int_{h_e}^{h}\frac{3}{h^3}\,dh \approx \frac{3}{2h_e^2}, \tag{B 7}$$

in terms of the "truncated" equilibrium film thickness (De Gennes 1985)

$$h_e = a\sqrt{3\gamma/2S}, \tag{B 8}$$

where $S$ is the spreading coefficient. This leads (B 6) to

$$T\dot{\Sigma}_f \approx SU, \tag{B 9}$$

which is (3.9).

## Appendix C. Derivation of (5.9) using the WKB ansatz

As in Hervet & De Gennes (1984) and Eggers & Stone (2004), we use the WKB ansatz to determine the asymptotic solution to (5.8) in the precursor film region.

According to this approximation, we assume that the solution $\psi$ varies slowly with $\zeta$ in the form of

$$\psi = \exp\left(\int g(\zeta)\,dz\right). \tag{C 1}$$

We write $g$ as a series:

$$g = g_0 + g_1 + \cdots, \tag{C 2}$$

such that $g_0 \gg g_1 \gg \cdots$. Substituting (C 1) with (C 2) into (5.8) yields, to leading order,

$$g_0^3 = \frac{3g_0}{\phi_{-\infty}^4}, \tag{C 3}$$

that is,

$$g_0 = \frac{\sqrt{3}}{\phi_{-\infty}^2}. \tag{C 4}$$

Knowing that $g_0' \ll g_0$ contributes to the first order correction and is compatible to $g_1$, the first correction yields

$$g_0 g_0' + 3g_0 g_1 = \frac{3g_1}{\phi_{-\infty}^4}. \tag{C 5}$$

We can then solve for $g_1$ by substituting $g_0$, given by (C 4), into (C 5), which yields

$$g_1 = -\frac{3}{2} (\ln g_0)'. \tag{C 6}$$

Finally, we use the first two leading order terms of $g$, by combining (C 4) and (C 6) in (C 2), to obtain the following expression for $\psi$, (C 1), as given by (5.9*a*):

$$\psi = g_0^{-3/2} \exp\left(\int g_0 \, d\zeta\right), \tag{C 7}$$

where

$$\int g_0 \, d\zeta = -\frac{\sqrt{3}(2-n)}{C^2(4-n)} (-\zeta)^{(4-n)/(2-n)}, \tag{C 8}$$

with $C = [3/(2-n)]^{1/(2-n)}$.